**Research article**

Philipp Grimm, Gary Razinskas, Jer-Shing Huang*, and Bert Hecht*

# Driving plasmonic nanoantennas at perfect impedance matching using generalized coherent perfect absorption

**Abstract:** Coherent perfect absorption (CPA) describes the absence of all outgoing modes from a lossy resonator, driven by lossless incoming modes. Here, we show that for nanoresonators that also exhibit radiative losses, e.g. plasmonic nanoantennas, a generalized version of CPA (gCPA) can be applied. In gCPA outgoing modes are suppressed only for a subset of (guided plasmonic) modes while other (radiative) modes are treated as additional loss channels - a situation typically referred to as perfect impedance matching. Here we make use of gCPA to show how to achieve perfect impedance matching between a single nanowire plasmonic waveguide and a plasmonic nanoantenna. Antennas with both radiant and subradiant characteristics are considered. We further demonstrate potential applications in background-free sensing.

*Corresponding authors: Bert Hecht, Nano-Optics & Biophotonics group, Department of Experimental Physics 5 & Röntgen Research Center for Complex Material Research (RCCM), Physics Institute, University of Würzburg, Am Hubland, 97074 Würzburg, Germany; and Jer-Shing Huang, Leibniz Institute of Photonic Technology, Albert-Einstein-Str. 9, 07754 Jena, Germany; Institute of Physical Chemistry and Abbe Center of Photonics, Friedrich-Schiller-Universität Jena, 07743 Jena, Germany; Research Center for Applied Sciences, Academia Sinica, 128 Sec. 2, Academia Road, Nankang District, 11529 Taipei, Taiwan; and Department of Electrophysics, National Chiao Tung University, Hsinchu 30010, Taiwan, E-mail: hecht@physik.uni-wuerzburg.de (B. Hecht), Jer-Shing.Huang@leibniz-ipht.de (J.-S. Huang)
**Philipp Grimm**, Nano-Optics & Biophotonics group, Department of Experimental Physics 5 & Röntgen Research Center for Complex Material Research (RCCM), Physics Institute, University of Würzburg, Am Hubland, 97074 Würzburg, Germany, E-mail: pgrimm@physik.uni-wuerzburg.de
**Gary Razinskas**, Nano-Optics & Biophotonics group, Department of Experimental Physics 5 & Röntgen Research Center for Complex Material Research (RCCM), Physics Institute, University of Würzburg, Am Hubland, 97074 Würzburg, Germany, E-mail: razinskas_g@ukw.de

## 1 Introduction

Perfect absorption of selected frequencies of coherent light is a special condition that can occur in systems that scatter light and possess lossy resonances. Such coherent perfect absorption (CPA) is based on complete destructive interference of all outgoing modes and corresponds to the time-reversed process of lasing at threshold [1-6]. At the CPA condition incoming modes are completely absorbed by the system and completely converted to other forms of energy, usually heat. Naturally, CPA occurs at Fabry-Pérot resonances of the lossy resonator. This is because the multiple reflections which accompany such resonances can lead to perfect destructive interference between the first reflected wave and all subsequent outgoing waves. This requires matching both their total phase and amplitude. Technically, a CPA condition corresponds to a zero eigenvalue of the scattering matrix associated with phase singularities located on the emitted mode's dispersion characteristic in the complex wavevector plane.

Being lossy is a rather natural property of plasmonic systems, which are therefore perfectly suited to exhibit CPA. Indeed, CPA has previously been proposed or observed in systems combining localized plasmonic resonators with photonic modes in dielectrics or free space [7-15]. Compared to dielectric Fabry-Pérot-type resonators originally used for CPA, plasmonic resonators, i.e. optical antennas, exhibit much more flexibility because they offer a variety of resonances whose properties can be tailored to exhibit e.g. radiant or sub-radiant characteristics. In any case, radiative losses cannot be ignored and even for sub-radiant modes contribute significantly to the overall losses of such an antenna with respect to the driving guided modes. Yet, in the original concept of CPA radiation losses are usually neglected.

Here we show that the concept of CPA, which requires at least one eigenvalue of the complete scattering matrix to vanish, can be generalized to situations where this applies only to a submatrix. As an example, we consider a semi-infinite single-mode gold plasmonic nanowire which is coupled to a single gold nanorod antenna via a gap. At least one zero eigenvalue of the scattering matrix is required only for the guided wire modes, while the infinite number of radiative modes of the nanorod are treated as losses. By doing so, we sacrifice the instructive, but practically not very relevant correspondence of CPA to the time-reversed version of lasing at threshold. On the other hand, generalized CPA (gCPA) can now be applied to any resonant absorber with both radiative and non-radiative loss channels, possibly including the absorption of light by quantum emitters. In the case of a single nanorod, studied here, obviously, the condition of gCPA corresponds to perfect impedance matching between the nanorod antenna and the nanowire transmission line since the incoming plasmonic mode is perfectly absorbed without any reflection [16-19]. Achieving perfect impedance matching is intrinsically challenging in plasmonic nanostructures due to the mismatch between the wavelength of guided plasmons and free space waves [20]. Depending on the chosen nanorod resonance we achieve under perfect impedance matching either strong far-field radiation or generation of heat.

In the proposed gCPA, the choice of the subspace in the output vector is flexible, depending on the application of interest. The subspace treatment is based on the fact that a global scattering matrix (S-matrix) may use a basis of guided modes and radiative modes. In the traditional formulation of CPA, only waveguide modes were considered and thus the S-matrix contained the coupling between guided input and output modes only. For the settings explored in the work of Stone [1], this ignoring of possible radiative modes is fully justified, whereas in the case of nanoplasmonics it is not. In this sense, our approach is a generalization of CPA. When the guided plasmon has zero reflection at the end of the wire we call it gCPA because the analysis of this effect follows the logic of Stone's CPA. Yet, the energy fed to the nanorod is not completely dissipated to heat, but some radiation occurs such that guided modes only are insufficient to capture the complete behavior of the system. While the example discussed here is specific, the underlying concept is more general than the original CPA. Recently, Sweeny and Stone reported a generalized theory of reflectionless scattering modes (RSM) [5, 6], which allows for reaching zero reflection of the selected input modes by evaluating the eigenvalue of the corresponding subset of the scattering matrix. The re-radiation from the system can be considered as one of the complementary output channels. While the RSM theory may also be applied to address resonators with radiative loss, the gCPA presented in this work was developed independently to address the coupling of a guided sub-wavelength surface plasmon mode to an optical nanoantenna. The underlying physics responsible for the zero reflection in this deep subwavelength system is clearly explained. We demonstrate the possibility of background-free sensing using a disturbed gCPA condition and coherent control of the radiation from a nanorod. Our findings establish gCPA as a tool in plasmonic nanocircuitry and nanoantenna design and technology.

## 2 Concept of generalized coherent perfect absorption

With dielectric waveguides and resonators, such as photonic waveguides feeding into a lossy cavity, the only dissipation mechanism in CPA is heat generation inside the cavity. Far-field radiation is not considered. Hence, a scattering matrix consisting of the coupling coefficients between the incoming and outgoing guided modes fully captures the modal conversion and energy exchange within and between the resonator and waveguides. However, the situation is different when it comes to plasmonic nanoantennas, where the oscillating surface plasmon leads to radiative decay in addition to the nonradiative loss into heat. In this case, the concept of gCPA would allow us to concentrate on the perfect absorption of the input mode of interest without fulfilling the requirements of CPA for the complete system. As an example, we illustrate the concept of gCPA with the case of a plasmonic nanoantenna driven by two semi-infinite single-mode plasmonic nanowires (Fig. 1). The resonator exhibits both Ohmic damping and far-field radiation. The latter can be expressed as a superposition of suitable free-space modes, e.g. plane waves propagating in different directions.

The total scattering matrix of the wire-rod-wire system in Fig. 1 is defined by

$$
\begin{pmatrix}
E_{guided,out,l} \\
E_{guided,out,r} \\
E_{rad,out,1} \\
E_{rad,out,2} \\
\vdots
\end{pmatrix}
= S_{global}
\begin{pmatrix}
E_{guided,in,l} \\
E_{guided,in,r} \\
E_{rad,in,1} \\
E_{rad,in,2} \\
\vdots
\end{pmatrix},
\tag{1}
$$

connecting *all* incoming with *all* outgoing modes. The four coefficients describing the coupling between the guided modes occupy the upper left corner of the scattering matrix as a two-by-two matrix,

$$
S_{global} =
\begin{pmatrix}
\begin{bmatrix} r & t \\ t & r \end{bmatrix} & \begin{matrix} c_{l1} & c_{l2} \\ c_{r1} & c_{r2} \end{matrix} & \cdots \\
\begin{matrix} c_{1l} & c_{1r} \\ c_{2l} & c_{2r} \end{matrix} & \begin{matrix} d_{11} & d_{12} \\ d_{21} & d_{22} \end{matrix} & \cdots \\
\vdots & \vdots & \ddots
\end{pmatrix}.
\tag{2}
$$

Here, $r$ and $t$ are the reflection and transmission coefficients of guided surface plasmons at the nanorod, the $c_{ij}$ are coefficients describing the modal coupling between surface plasmons and radiative modes, and $d_{ij}$ establish the coupling among radiative modes. In this generalized framework, the "complete" CPA requires to find a condition for which an eigenvalue of the *entire* scattering matrix becomes zero. This means *all* outgoing channels, guided modes and radiation, are turned to zero simultaneously. Achieving such complete CPA can be experimentally challenging or even impossible for resonators with radiative loss, such as nanoantennas.

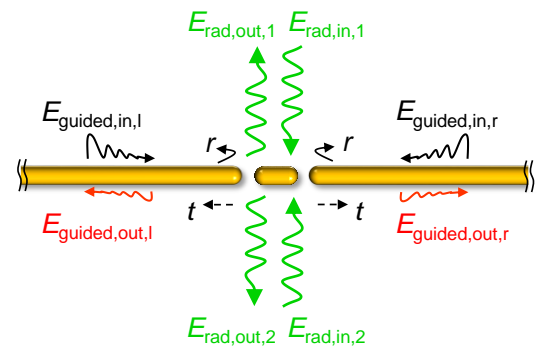

**Figure 1:** (a) Sketch of a symmetric wire-rod-wire system. In general, guided surface plasmons $E_{guided}$ as well as radiative modes $E_{rad}$ have to be used as incoming and outgoing modes. Within the generalized CPA formalism in this work, only the subset of guided surface plasmons are considered as incoming modes.

In this work, the input and output modes we are interested in are the left and right-propagating fundamental guided modes on a plasmonic nanowire. All other modes are treated as loss channels, including the radiation of the nanoantenna. The input field vector that is multiplied with the scattering matrix therefore only contains finite values for the top two components in Eq. (1). All the other components are set to zero. Now, gCPA requires to only find and zero the eigenvalue of the subset of the resulting upper left two-by-two matrix, as marked by the square brackets in Eq. (2), which describes the coupling between the guided modes only. Since the major part of the scattering matrix which connects the radiative modes is not considered in the required diagonalization the coupling to radiative modes will in general not be zero, meaning that complete CPA of the whole system is in general not achieved. The coupling to radiative modes is treated as a part of the loss of the guided plasmonic modes. As we will show in the following, by choosing the nanoantenna to be subradiant or superradiant, the amount of radiation losses can be controlled to a large degree. Generalized CPA, i.e. a zero eigenvalue of a sub-matrix, occurs if the energy from a few incoming guided wire plasmons is entirely converted to other forms of energy, no matter radiative or nonradiative, featuring the zero reflection of the input modes. This provides a practical route to find the condition for perfect impedance matching between plasmonic waveguides and plasmonic nanoantennas in nanocircuitry.

# 3 Perfectly impedance-matched nano-antenna using generalized coherent perfect absorption

In the following, we demonstrate that the concept of gCPA allows us to find the condition for perfect impedance matching between a single wire plasmonic waveguide and nanorod optical antenna. At this condition, the guided wire plasmons are completely injected into the antenna without any reflection and the injected power is branched into radiative and non-radiative loss, depending on the antenna mode and excitation conditions.

The system under study consists of a semi-infinite gold nanowire with a circular cross section (30 nm diameter) in the vacuum in the near infrared regime (310 to 400 THz, corresponding to vacuum wavelength 749~967 nm). The wire is terminated by a hemispherical end cap. A gold nanorod with the same diameter and end caps is coupled to the nanowire over a few nanometer wide gap (see Fig. 2a). Throughout this work, the minimum gap width studied is 3 nm. Since all gaps are sufficiently large we neglect quantum effects. For sub-nanometer gaps, more sophis-ticated methods might be needed to obtain the fundamental reflection and transmission coefficients under the influence of quantum effects. As long as these coefficients could be obtained, the formalism would work in the same way. Due to the relatively small wire diameter compared to the vacuum wavelength, the nanowire supports only the fundamental plasmonic $TM_0$ eigenmode [21, 22]. Its transverse mode profile is displayed in the inset of Fig. 2a. The complex propagation constant

$$k = \beta + i\alpha \qquad (3)$$

is found using the finite-difference frequency-domain method (MODE Solutions, Lumerical Solutions Inc.) [23]. Here, $i$ is the imaginary unit, $\beta = 2\pi/\lambda_{eff}$ is the propagation constant with $\lambda_{eff}$ the effective wavelength, and $\alpha$ is the field decay constant due to Ohmic losses. The dielectric function of gold is modeled using single-crystal data [24]. The obtained eigenmode is used as a source in three-dimensional finite-difference time-domain simulations (FDTD Solutions, Lumerical Solutions Inc.). Taking plasmon reflection at the wire termination into account the electric near-field intensity distribution along the semi-infinite wire is [25, 26]

$$|E(x)|^2 = \left| E_0 \left[ e^{ikx} + e^{ik(x_0 - x)} \Gamma e^{ikx_0} \right] \right|^2, \qquad (4)$$

where $E_0$ is the initial amplitude of the mode, $x$ is the spatial coordinate in the propagation direction, $x_0$ is the distance between the mode source injection point at $x = 0$ and the end of the cylindrical part of the wire at $x = 2985$ nm, and $\Gamma$ is the complex reflection coefficient obtained by fitting the simulated standing wave pattern of the electric near-field intensity 5 nm away from the wire surface with Eq. (4). This yields $\Gamma$ as the only free parameter. Since the mode reflection is very sensitive to the exact condition of the wire termination [27, 28], the nanorod coupled via the nanosized gap can alter the reflection coefficient and the standing wave pattern drastically. Fig. 2b shows two distinctively different exemplary near-field standing wave patterns corresponding to the absence of a nanorod and a specifically chosen rod length to be discussed below. In absence of a vicinal nanorod a pronounced standing wave pattern is observed corresponding to a high reflectivity ($|\Gamma|$ = 95%). For a nanorod length of 346 nm, nearly zero reflectivity is observed, accompanied by strong energy localization on the nanorod. The absence of the reflection corresponds to perfect impedance matching between the nanowire (transmission line) and the nanorod (load) and suggests an effective scheme to drive the nanorod antenna via a single-wire transmission line (see Visualization 1 and section I of Supplementary Material). In optical nanocircuitry, the characteristic impedance is determined by the electromagnetic fields of the guided mode on the nanowire. The degree of impedance matching, described by the complex reflectivity $\Gamma$, can be evaluated by characterizing the standing waves of the optical near field around the nanowire waveguide termination. Once the characteristic impedance of the guided mode and the reflectivity at a given termination are known, the impedance of the load, i.e. the nanoantenna, can be calculated using the complex reflectivity $\Gamma$ [25] .

In the following, we show that the observed zero reflectivity corresponds to a gCPA condition of the guided plasmonic mode that can be achieved for both a superradiant and a subradiant resonance of the nanorod. This offers the opportunity to control the branching ratio of the absorbed power to either far-field radiation or absorption. We describe the overall reflectivity of the nano-wire termination using a semi-analytical approach, in which the gap-coupled nanorod is considered as an effective Fabry-Pérot resonator [29, 30]. The reflectivity of the nanorod termination can then be calculated by summing up an infinite geometrical series involving the open-end reflectivity and the reflection and transmission at a gap between two semi-infinite wires (see section II of Supplementary Material). These three parameters are determined numerically by means of FDTD simulations for a range of frequencies and gap widths as described in Supplementary Material. The resulting analytical expression allows us to instantaneously calculate the overall reflectivity of the wire-rod system. For further analysis, we note that the overall reflectivity also corresponds to the scattering `matrix' of the system for the subset of guided modes which is a scalar for the 1D single-mode case considered here. Perfect destructive interference, i.e. zero reflectivity, requires the directly reflected mode (green arrow, Fig. 2a) and the transmitted mode, consisting of the infinite sum of transmitted waves over the gap (dark red arrows, Fig. 2a), to have opposite phases and equal amplitudes. To achieve this, the nanorod must provide just the right amount of loss and the gap must be chosen correctly to transmit a just sufficient portion of the mode for each round trip. In this regard, the length-dependent Ohmic losses of the plasmonic waveguide indeed become important and beneficial to enable unidirectional nanoscale gCPA.

Fig. 2c shows the calculated reflectance (squared reflection amplitude, $|\Gamma|^2$) as a function of nanorod length at various frequencies for a fixed gap of 4 nm. Multiple local minima are obtained at different orders of nanorod resonances. Indeed, the second-order minimum reaches zero reflection, i.e. gCPA as we prove below. Interestingly, all the rod lengths for which reflection minima occur coincide with those of solitary nanorods showing a scattering resonance at that wavelength as if the proximity of the nanowire did not perturb the resonance of the nanorod. This coincidence marks a special "nonbonding" condition, where the phase of the electric field on opposing sides of the gap exhibits $\pi/2$ difference (see section III of Supplementary Material). gCPA must show up at the "nonbonding" resonant rod length because CPA means no reflection and thus no charge accumulation at the semi-infinite wire end as if the wire was continuous. The nonbonding condition with appropriate loss offers the necessary phase and amplitude for perfectly destructive interference of the reflected mode. Another interesting feature is that gCPA is not associated with specific resonance orders. For example, at 440 THz, the smallest reflection occurs at the first-order resonance of the nanorod. At 360 THz (see Fig. 2d), the nanorod's second-order resonance leads to the lowest reflection intensity. This suggests the flexibility to achieve gCPA with even and odd resonance orders and thus the possibility to select the loss mechanisms, as will be discussed later.

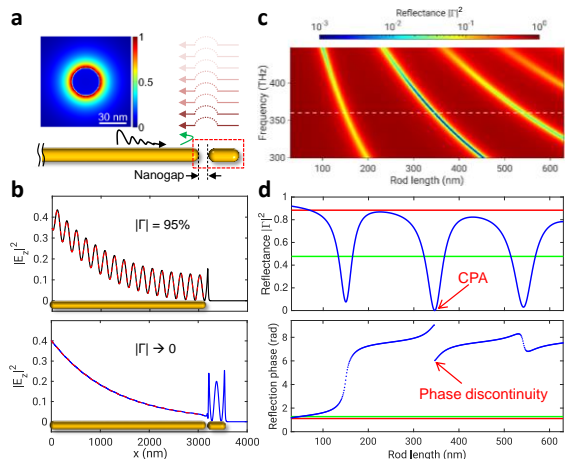

**Figure 2:** (a) Reflection of the right propagating guided plasmonic mode (black arrow) at the termination of a gold nanowire. The total mode reflection at the wire termination (red dashed rectangle) is determined by the interference between the directly reflected mode at the gap (green arrow) and a series of coherent transmissions from the mode oscillating on the nanorod (dark red arrows). See section II of Supplementary Material for details. Inset: modal profile of the guided wire mode. (b) Simulated standing wave patterns of the guided wire mode (electric near-field intensity) for a termination open to vacuum (upper panel) and facing a 346-nm nanorod (lower panel) via a 5-nm gap. The red dashed lines show the fits of Eq. (4) to these standing wave patterns (FDTD) using the complex reflectivity $\Gamma$ as the only free fit parameter. (c) Calculated reflectance ($|\Gamma|^2$) as a function of the nanorod length and frequency. The gap is 4 nm. (d) Reflection intensity (upper panel) and phase (lower panel) as a function of the nanorod length at 360 THz, marked by the white dashed line in (c). The red and green horizontal lines mark the reflectance and phase of a termination open to vacuum and a gap in an infinitely long wire, respectively.

To distinguish good absorption (local reflection minima) from perfect absorption (gCPA), we investigate the continuity of the reflection phase. Fig. 2d displays the reflectance and the reflection phase at 360 THz as a function of the nanorod length. The first, partial reflection dip, observed at nanorod length = 153 nm, exhibits a narrow pseudo-Lorentzian line shape. The corresponding phase transition is steep but still continuous. The second reflection minimum at a rod length of 346 nm (see Fig. 2b, lower panel) approaches zero, i.e. truly perfect absorption. The corresponding phase changes discontinuously. We would like to emphasize that this phase jump does not originate from an irrelevant $2\pi$ jump. It is caused by the vortex-like phase singularity (Fig. 3a) which accompanies the zero reflectance of the gCPA state. A meaningless $2\pi$ jump would not give rise to such a singularity. The difference between a local minimum and a truly zero reflection due to CPA is that true CPA happens when the phase singularity hits the dispersion curve of the guided mode. If the phase singularity is close but does not exactly reside on the dispersion curve of the guided mode, the reflectivity would just be a local minimum with a smooth transition of phase. In numerical simulations, a finite reflectivity is inevitable due to limited finite mesh size and frequency step used in the simulations. In reality, the resolution is limited by the diameter of an atom (the smallest spatial step) and the frequency bandwidth of the source. The narrow bandwidth of the gCPA dip implies that ultrashort pulsed excitation with broad bandwidth is not compatible with gCPA.

## 4 Shifting the phase singularity on the complex wave vector plane

In order to prove that we indeed observe a generalized version of CPA, in Fig. 3 we plot the reflectance and reflection phase over the complex wavevector plane [1, 31] since the attenuation constant $\alpha$ of the guided plasmonic $TM_0$ eigenmode may in principle be tuned by adding gain or additional absorption losses. For the selected range of frequencies, the reflectivity displays a zero and a pole of its amplitude accompanied by phase singularities in the positive and negative $\alpha$ range, respectively. Further poles and zeros exist for each order of Fabry-Pérot resonances of the nanorod (not shown). The plasmonic mode dispersion relation can be used to derive a relation between the propagation constant $\beta$ and the damping $\alpha$ of the guided $TM_0$ mode (see section IV of Supplementary Material). This relation is also plotted as a white solid line (SP mode) into the same complex wavevector plane. To achieve zero reflection, the phase singularity corresponding to zero reflectivity must reside on this curve describing allowed $\alpha$ and $\beta$ combinations of the mode [32], like the case displayed in Fig. 3a. By changing the geometry of the termination, such as the gap size and nanorod length, the position of the phase singularities can be moved freely within the complex wavevector plane.

Fig. 3b shows the trajectories of the phase singularities of the 1st- and 2nd-order resonances of the nanorod upon increasing the gap size and the rod length. While increasing the nanorod length shifts the phase singularities towards lower propagation constant, increasing the gap size mainly moves the singularity towards smaller attenuation constants. Within the experimentally accessible range of geometrical parameters, multiple singularities corresponding to different Fabry-Pérot resonance orders of the nanorod are available. This allows us to rationally design a wire-rod system to achieve gCPA using super-radiant or sub-radiant nanorod resonances, offering the opportunity to choose the dissipation mechanism for the absorbed energy. Thus, the gCPA analysis on the complex wavevector plane serves as a theoretical design tool for perfect impedance matching. The transfer matrix calculations yield fast results and avoid time-consuming numerical parameter sweeps. The good agreement between semi-analytical and purely numeric results for reflection amplitude and phase is demonstrated in Fig. S2 of the Supplementary Material. A movie showing the motion of the phase singularities and the crossing of the mode dispersion, accompanied by the reflectance spectrum, is provided in Visualization 2 (see section V of Supplementary Material for video description). Another important feature seen in Fig. 3 is the phase singularity point in the lower half of the complex plane, where the attenuation constant $\alpha$ is negative. This phase singularity is associated with a reflectivity pole. For negative $\alpha$ the guided mode is amplified, suggesting the possibility of surface plasmon amplification by stimulated emission [9, 33, 34].

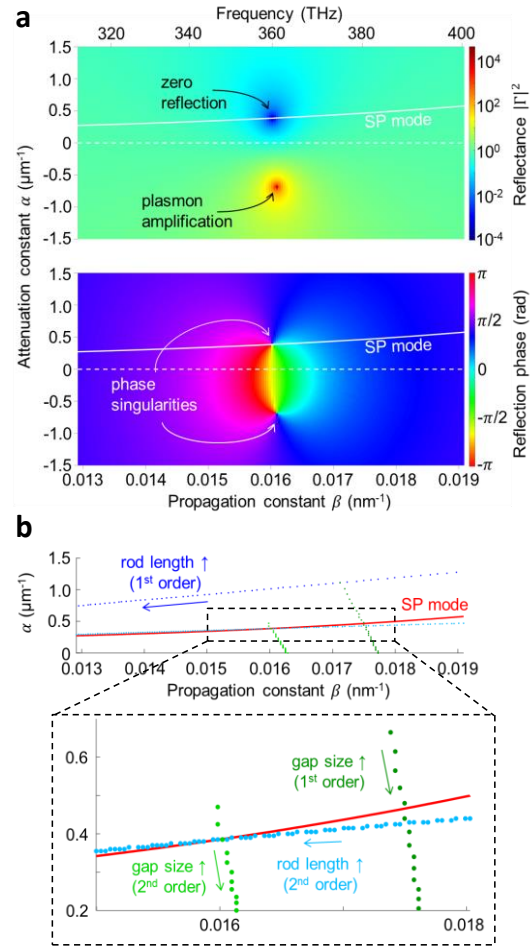

**Figure 3:** (a) The reflectance (upper panel) and reflection phase (lower panel) plotted over the complex wavevector plane for a specific terminal condition (rod length = 346 nm, gap = 4 nm), at which one of the phase singularities hits the dispersion curve of the guided mode, leading to zero reflection for one specific value of $\beta$. $\beta$ and $\alpha$ are the real and imaginary parts of the wavevector. Negative $\alpha$ represents gain. (b) Upper panel: trajectories of the reflection phase singularities of the first- and second-order nanorod resonance upon sweeping the gap size (increment 0.5 nm per dot) and nanorod length (increment 1 nm). Dark blue and dark green dots mark the trajectories of the phase singularity of the first-order resonance upon increasing the rod length (126 nm to 187 nm) and gap size (4 nm to 19 nm), respectively. Light blue and light green dots mark the trajectories of the phase singularity of the second-order resonance upon sweeping the rod length (290 nm to 431 nm) and gap size (3 nm to 20 nm), respectively. The red solid curve depicts the allowed $\alpha$ and $\beta$ for the $TM_0$ mode on the nanowire. Lower panel: enlarged plot corresponding to the area marked by the dashed rectangle in the upper panel.

# 5 Exemplary applications of generalized coherent perfect absorption

The gCPA approach developed in this work offers a broad range of applications. In principle, it can be applied to study the coupling of any guided mode into lossy cavities where the total loss may include radiative channels. In particular, structures consisting of discrete elements are straightforward to treat as they are composed of elementary reflection and transmission events. In the following, we will discuss two applications of gCPA in plasmonic nanocircuitry, (i) driving bright or dark nanoantenna modes with perfectly matched impedance and (ii) background-free nanoscale sensing. Fig. 4a shows the efficiency of radiative and non-radiative loss of a unidirectional gCPA-driven single nanorod in the first and second-order resonance. The same quantities of the open end of a semi-infinite nanowire are also plotted for reference. According to the symmetry of the currents on the nanorod, the first- and second-order resonances are super-radiant and sub-radiant, respectively. Therefore, gCPA based on the first-order resonance (rod length = 157 nm, gap = 9 nm) results in 50% far-field radiation efficiency, whereas CPA achieved with the second-order resonance (rod length = 346 nm, gap = 5 nm) leads to 85% non-radiative energy dissipation into heat. Obviously, these values could be further improved by optimized antenna designs.

Two-port CPA using photonic modes with dielectric or plasmonic resonators has been used to demonstrate coherent control of light with light without using nonlinear effects [2, 7, 13, 14]. Guided plasmons on two plasmonic nanowires sandwiching one nanorod resonator can also be used to realize two-port gCPA [35] (see section VI of

Supplementary Material). With gCPA based on a super-radiant resonance mode, the relative phase of the two plasmon inputs can be used to coherently control the radiation of the plasmonic nanorod, i.e. to control the emission of a plasmon-driven transmitting optical nanoantenna by surface plasmon without using a nonlinear process.

For sensing applications, we use the sub-radiant nanorod resonance (rod length 346 nm, gap 5 nm) to achieve gCPA. Tiny perturbations of the nanorod's terminal conditions will shift the phase singularity in the complex plane and lead to the destruction of gCPA. As a result, a finite reflection against a completely dark background can be detected with high signal-to-background contrast, similar to the case of dark-field scattering or single-molecule fluorescence detection. Assuming that the major source of noise in an experiment is shot noise, which is proportional to the square root of the signal intensity, the signal-to-noise (S/N) ratio is proportional to the amplitude of the reflection. It should be noted, though, that the amplitude cannot be increased arbitrarily by increasing the input intensity due to local heating of the nanorod. In Fig. 4b, we compare the change of the reflection amplitude upon local perturbations of a wire-rod system with that of a bare single wire probe. A perturbation is introduced by approaching a tiny dielectric glass nanosphere (diameter = 20 nm) to the termination (inset of Fig. 4b). The wire-rod system used here shows gCPA based on the sub-radiant 2nd-order resonance, i.e. the 2nd dip in Fig. 2d. There is a rapid increase of the reflection amplitude as the silica nanosphere approaches the nanorod along the wire axis to a distance less than 20 nm. A pronounced nonlinear increase in the reflection amplitude is obtained when the separation is below 10 nm, showing ultimate sensitivity to perturbations in close vicinity of the probe. The analytically predicted reflectivity increase of the wire-rod system is perfectly reproduced by FDTD simulations (data not shown). We conceive that even attachment or detachment of single proteins should be detectable as a notable increase in reflectivity [36-38].

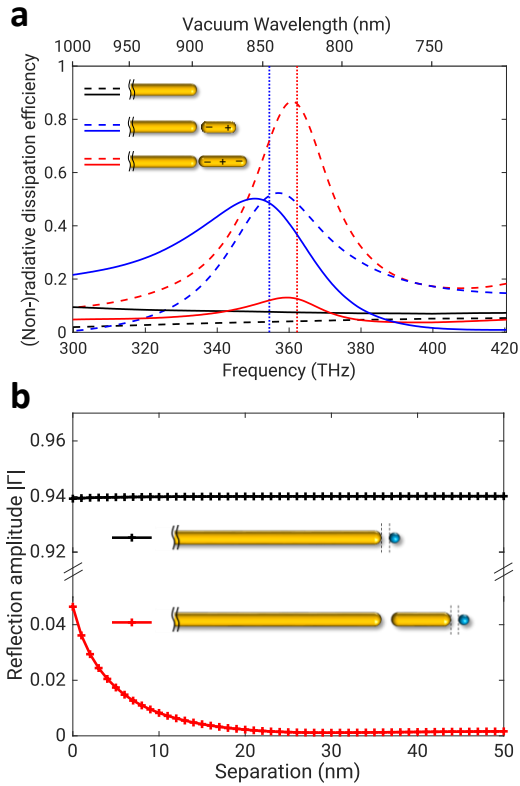

**Figure 4:** (a) Radiative (solid lines) and non-radiative (dashed lines) dissipation efficiency of gCPA-driven nanorod at the first- (blue) and second-order resonance (red). The radiative and nonradiative dissipation efficiencies of the open end of a semi-infinite wire is also plotted for reference (black). The vertical dotted lines indicate the gCPA frequencies for first and second order resonances. Legends also include the charge distribution on the nanorods for first- and second-order resonances. (b) Comparison of the reflection amplitude to a local perturbation (approaching dielectric sphere, diameter 20 nm) of a simple wire sensor (black) compared to a gCPA wire-rod sensor (red). For the latter, a pronounced change in reflection amplitude is observed for separations below 10-20 nm.

## 6 Conclusion

In conclusion, at the example of a semi-infinite plasmonic nanowire terminated by a gold nanorod, we present a new type of nanoscale, near-field energy transfer and perfect absorption based on a generalized CPA concept. We demonstrate that tuning the geometry of the nanorod can be used to deterministically move the reflectivity phase singularity in the complex wavevector plane. This allows us to find zero eigenvalue conditions for the selected subset of guided plasmonic modes of the scattering matrix. Since for gCPA no reflection occurs, the condition also corresponds to perfect impedance matching of the guided surface plasmon with respect to the nanorod antenna. The gCPA condition completely cloaks the wire termination and virtually decouples the nanorod from the feeding structure. We further demonstrated that the gCPA condition is very sensitive to changes in the local environment and thus useful for background-free ultrasensing. A correctly designed nanowire-rod system could also serve as a probe for near-field sensing. Furthermore, we demonstrate the coherent control of nanoantenna radiation using two-port gCPA. Using the concept of gCPA is necessary for systems in which radiative losses may occur, e.g. plasmonic nanoresonators and quantum emitters. The generalization causes a breakdown of the equivalence between time-reversed lasing at threshold and CPA, although a SPASER condition (reflectivity pole) is still predicted if the loss on the nanorod is turned into gain.

**Acknowledgements:** We thank Tobias Helbig and Tobias Hofmann for pointing to the transfer matrix formalism, as well as Thorsten Feichtner and René Kullock for fruitful discussions. The authors acknowledge the financial support from the DFG via grants HU2626/3-1(423427290), HU2626/6-1 (447515653), HU2626/5-1(445415315), SFB 1375 NOA (398816777) and HE5618/6-1.

**Conflict of interest statement:** The authors declare no conflicts of interest.

## References

1. Y. D. Chong, L. Ge, H. Cao, and A. D. Stone, "Coherent perfect absorbers: time-reversed lasers," Phys. Rev. Lett. **105**, 053901 (2010).

2. W. Wan, Y. Chong, L. Ge, H. Noh, A. D. Stone, and H. Cao, "Time-Reversed Lasing and Interferometric Control of Absorption," Science **331**, 889-892 (2011).

3. D. G. Baranov, A. Krasnok, T. Shegai, A. Alù, and Y. Chong, "Coherent perfect absorbers: linear control of light with light," Nat. Rev. Mater. **2**, 17064 (2017).

4. K. Pichler, M. Kühmayer, J. Böhm, A. Brandstötter, P. Ambichl, U. Kuhl, and S. Rotter, "Random anti-lasing through coherent perfect absorption in a disordered medium," Nature **567**, 351-355 (2019).

5. W. R. Sweeney, C. W. Hsu, and A. D. Stone, "Theory of reflectionless scattering modes," Physical Review A **102**, 063511 (2020).

6. A. D. Stone, W. R. Sweeney, C. W. Hsu, K. Wisal, and Z. Wang, "Reflectionless excitation of arbitrary photonic structures: a general theory," Nanophotonics **10**, 343-360 (2021).

7. J. Zhang, K. F. MacDonald, and N. I. Zheludev, "Controlling light-with-light without nonlinearity," Light Sci. Appl. **1**, e18-e18 (2012).

8. S. Dutta-Gupta, R. Deshmukh, A. Venu Gopal, O. J. F. Martin, and S. Dutta Gupta, "Coherent perfect absorption

mediated anomalous reflection and refraction," Opt. Lett. **37**, 4452-4454 (2012).

9. H. Noh, Y. Chong, A. D. Stone, and H. Cao, "Perfect coupling of light to surface plasmons by coherent absorption," Phys. Rev. Lett. **108**, 186805 (2012).

10. J. W. Yoon, G. M. Koh, S. H. Song, and R. Magnusson, "Measurement and Modeling of a Complete Optical Absorption and Scattering by Coherent Surface Plasmon-Polariton Excitation Using a Silver Thin-Film Grating," Phys. Rev. Lett. **109**, 257402 (2012).

11. R. Bruck, and O. L. Muskens, "Plasmonic nanoantennas as integrated coherent perfect absorbers on SOI waveguides for modulators and all-optical switches," Opt. Express **21**, 27652-27661 (2013).

12. M. J. Jung, C. Han, J. W. Yoon, and S. H. Song, "Temperature and gain tuning of plasmonic coherent perfect absorbers," Opt. Express **23**, 19837-19845 (2015).

13. T. Roger, S. Vezzoli, E. Bolduc, J. Valente, J. J. F. Heitz, J. Jeffers, C. Soci, J. Leach, C. Couteau, N. I. Zheludev, and D. Faccio, "Coherent perfect absorption in deeply subwavelength films in the single-photon regime," Nat. Commun. **6**, 7031 (2015).

14. X. Fang, K. F. MacDonald, and N. I. Zheludev, "Controlling light with light using coherent metadevices: all-optical transistor, summator and invertor," Light Sci. Appl. **4**, e292-e292 (2015).

15. R. Alaee, Y. Vaddi, and R. W. Boyd, "Dynamic coherent perfect absorption in nonlinear metasurfaces," Optics Letters **45**, 6414-6417 (2020).

16. N. Engheta, A. Salandrino, and A. Alù, "Circuit Elements at Optical Frequencies: Nanoinductors, Nanocapacitors, and Nanoresistors," Phys. Rev. Lett. **95**, 095504 (2005).

17. A. Alù, and N. Engheta, "Input Impedance, Nanocircuit Loading, and Radiation Tuning of Optical Nanoantennas," Phys. Rev. Lett. **101**, 043901 (2008).

18. J.-J. Greffet, M. Laroche, and F. Marquier, "Impedance of a Nanoantenna and a Single Quantum Emitter," Phys. Rev. Lett. **105**, 117701 (2010).

19. J.-S. Huang, J. Kern, P. Geisler, P. Weinmann, M. Kamp, A. Forchel, P. Biagioni, and B. Hecht, "Mode Imaging and Selection in Strongly Coupled Nanoantennas," Nano Lett. **10**, 2105-2110 (2010).

20. L. Novotny, "Effective wavelength scaling for optical antennas," Phys. Rev. Lett. **98**, 266802 (2007).

21. S. Zhang, H. Wei, K. Bao, U. Hakanson, N. J. Halas, P. Nordlander, and H. Xu, "Chiral surface plasmon polaritons on metallic nanowires," Phys. Rev. Lett. **107**, 096801 (2011).

22. C. Rewitz, T. Keitzl, P. Tuchscherer, J. S. Huang, P. Geisler, G. Razinskas, B. Hecht, and T. Brixner, "Ultrafast plasmon propagation in nanowires characterized by far-field spectral interferometry," Nano Lett. **12**, 45-49 (2012).

23. Z. Zhu, and T. G. Brown, "Full-vectorial finite-difference analysis of microstructured optical fibers," Opt. Expr. **10**, 853-864 (2002).

24. R. L. Olmon, B. Slovick, T. W. Johnson, D. Shelton, S.-H. Oh, G. D. Boreman, and M. B. Raschke, "Optical dielectric function of gold," Phys. Rev. B **86**, 235147 (2012).

25. J.-S. Huang, T. Feichtner, P. Biagioni, and B. Hecht, "Impedance Matching and Emission Properties of Nanoantennas in an Optical Nanocircuit," Nano Lett. **9**, 1897-1902 (2009).

26. G. Razinskas, P. Biagioni, and B. Hecht, "Limits of Kirchhoff's Laws in Plasmonics," Sci. Rep. **8**, 1921 (2018).

27. E. S. Barnard, J. S. White, A. Chandran, and M. L. Brongersma, "Spectral properties of plasmonic resonator antennas," Optics Express **16**, 16529-16537 (2008).

28. E. Feigenbaum, and M. Orenstein, "Ultrasmall volume plasmons, yet with complete retardation effects," Phys. Rev. Lett. **101**, 163902 (2008).

29. J. Dorfmüller, R. Vogelgesang, R. T. Weitz, C. Rockstuhl, C. Etrich, T. Pertsch, F. Lederer, and K. Kern, "Fabry-Pérot Resonances in One-Dimensional Plasmonic Nanostructures," Nano Lett. **9**, 2372-2377 (2009).

30. T. H. Taminiau, F. D. Stefani, and N. F. van Hulst, "Optical Nanorod Antennas Modeled as Cavities for Dipolar Emitters: Evolution of Sub- and Super-Radiant Modes," Nano Lett. **11**, 1020-1024 (2011).

31. E. S. Wegert, Gunter, "Phase Plots of Complex Functions: A Journey in Illustration," Notices Amer. Math. Soc. **58**, 768-780 (2011).

32. A. Krasnok, D. Baranov, H. Li, M.-A. Miri, F. Monticone, and A. Alú, "Anomalies in light scattering," Adv. Opt. Photon. **11**, 892-951 (2019).

33. R. F. Oulton, V. J. Sorger, T. Zentgraf, R.-M. Ma, C. Gladden, L. Dai, G. Bartal, and X. Zhang, "Plasmon lasers at deep subwavelength scale," Nature **461**, 629-632 (2009).

34. Y.-J. Lu, J. Kim, H.-Y. Chen, C. Wu, N. Dabidian, C. E. Sanders, C.-Y. Wang, M.-Y. Lu, B.-H. Li, X. Qiu, W.-H. Chang, L.-J. Chen, G. Shvets, C.-K. Shih, and S. Gwo, "Plasmonic Nanolaser Using Epitaxially Grown Silver Film," Science **337**, 450 (2012).

35. H. Park, S.-Y. Lee, J. Kim, B. Lee, and H. Kim, "Near-infrared coherent perfect absorption in plasmonic metal-insulator-metal waveguide," Opt. Express **23**, 24464-24474 (2015).

36. I. Ament, J. Prasad, A. Henkel, S. Schmachtel, and C. Sonnichsen, "Single unlabeled protein detection on individual plasmonic nanoparticles," Nano Lett. **12**, 1092-1095 (2012).

37. P. Zijlstra, P. M. Paulo, and M. Orrit, "Optical detection of single non-absorbing molecules using the surface plasmon resonance of a gold nanorod," Nat. Nanotech. **7**, 379-382 (2012).

38. V. R. Dantham, S. Holler, C. Barbre, D. Keng, V. Kolchenko, and S. Arnold, "Label-free detection of single protein using a nanoplasmonic-photonic hybrid microcavity," Nano Lett. **13**, 3347-3351 (2013).

Philipp Grimm, Gary Razinskas, Jer-Shing Huang*, and Bert Hecht*

# Driving plasmonic nanoantennas at perfect impedance matching using generalized coherent perfect absorption

## Supplementary Material

## Contents

# I. Movie of perfect near-field driving of plasmonic nanoantenna modes

This movie is available online.

Description:
The movie displays the time evolution of the electric near-field distribution ($E_z$) of an open-end nanowire (frequency 362.2 THz), a wire-rod system with gCPA at the first-order antenna resonance (rod length 157 nm, gap width 9 nm, frequency 354.5 THz), and a wire-rod system with gCPA at the second-order antenna resonance (rod length 346 nm, gap width 5 nm, frequency 362.2 THz) in the $xz$-plane. A standing wave pattern arises in the open-end case as a result of high reflection of the guided wire plasmon. In contrast, both antenna modes are resonantly fed by the right-propagating wire plasmon if a nanorod is attached. The neat combination of correct rod length, gap size, and frequency ensures perfect impedance matching, *i.e.* the back-reflection into the semi-infinite wire is completely suppressed.

# II. Analytical model for the reflectivity: The transfer matrix algorithm

In this section, we model the wire plasmon reflectivity at a single nanorod attached to a semi-infinite wire via a small gap. The model is based on a semianalytical approach: It is demonstrated that the reflectivity of any attached nanorod can be obtained by only considering reflections at the open end of an infinite wire, reflections at a gap in an infinite wire, as well as the transmission of a gap in an infinite wire. To obtain the reflectivity of a nanorod, we consider the interference effects of the direct reflection at the nanogap and the fields which are fed back to the wire across the gap after having undergone an infinite number of Fabry-Pérot oscillations on the nanorod (Fig. S1(a)).

The superposition of these contributions result in the total reflection $\Gamma$,

$$\Gamma = R_g + \frac{T_g^2}{R_g} \cdot \sum_{k=1}^{n} \left[ R_e R_g e^{2(-\alpha+i\beta)L_{net}} \right]^k. \qquad \text{(Eq. S1)}$$

Here, $R_g$ and $T_g$ are reflectivity at and transmission across the gap in an infinitely long wire, respectively. $R_e$ denotes the reflectivity at the end of a semi-infinite wire open to vacuum. All quantities also include radiation losses which is essential to find generalized CPA conditions. The guided $TM_0$ eigenmode is characterized by its propagation constant $\beta$ and attenuation constant $\alpha$. These quantities are determined by numerical simulations using the finite-difference frequency-domain and finite-difference time-domain method (MODE Solutions and FDTD Solutions, Lumerical Solutions Inc.). The complex coefficients $R_g$, $T_g$, and $R_e$ as functions of gap width and frequency are obtained by mode expansion calculations using Lumerical's port objects. The required electromagnetic field profiles are recorded at the end of each cylindrical section in the plane perpendicular to the wire axis (see vertical dashed lines in Figs. S1(a)-(b)). Perfectly matched layers (PML) are applied in all directions as boundaries of the simulation volume. The nanowire extends through the PML to mimic a semi-infinite geometry. $L_{net}$ is the net length of the nanorod, *i.e.* the length of its cylindrical part without the hemispherical endcaps and $n$ represents the number of oscillation roundtrips on the nanorod. We consider the case $n \to \infty$ and simplify the sum in (Eq. S1) using the limit of the geometric series:

$$\Gamma = R_g + \frac{T_g^2}{R_g} \cdot \left( \frac{1}{1 - R_e R_g e^{2(-\alpha+i\beta)L_{net}}} - 1 \right) \qquad \text{(Eq. S2)}$$

The same result for $\Gamma$ is obtained using a transfer matrix approach [1].

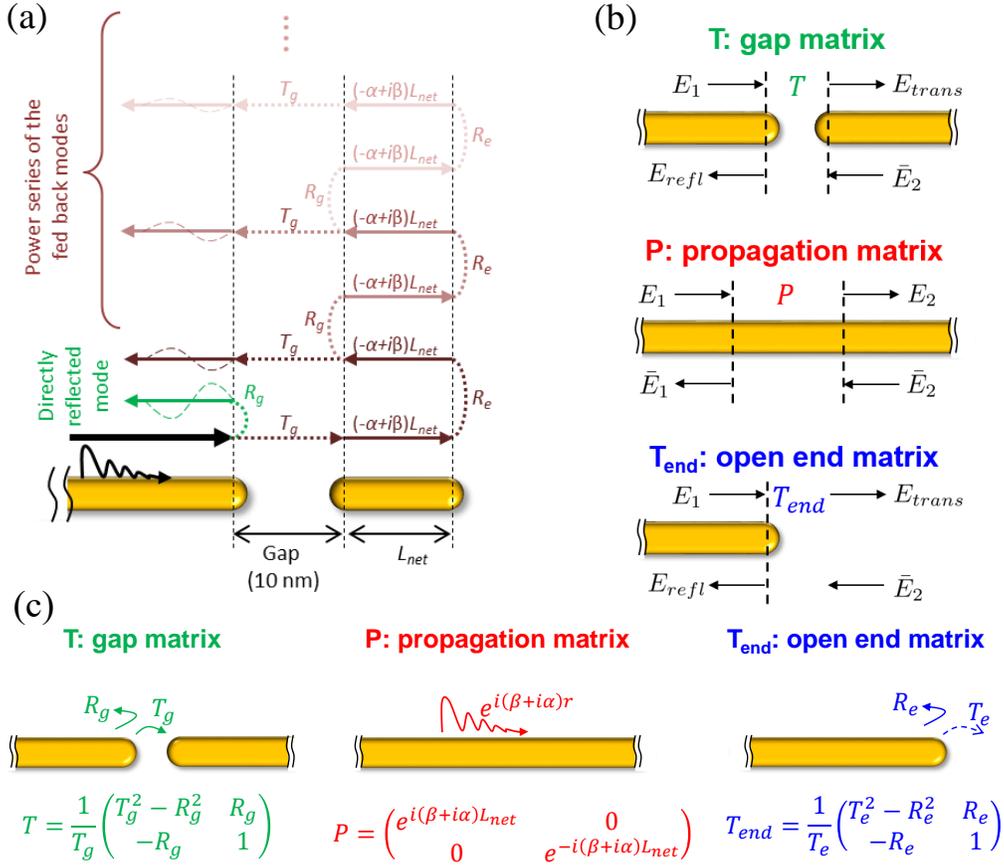

**Figure S1.** (a) Schematic illustration of the interference between the direct reflection and the power series of the fed back modes from the nanorod. (b) Definition of input and output quantities for the transfer matrices $T$ (reflection/transmission at gap), $P$ (propagation along cylindrical section), and $T_{end}$ (reflection/transmission at open end). (c) Schematic illustrations of the gap matrix ($T$), propagation matrix ($P$), and the open end matrix ($T_{end}$).

As explained in the main text, we choose the subset of incoming and outgoing wire plasmons as basis while radiation into free space is treated as loss mechanism. The transmission, reflection, and propagation properties of the gap, nanorod, and open end (see Fig. S1(b)) are modeled by the gap transmission matrix ($T$), propagation matrix ($P$), and the open end matrix ($T_{end}$), respectively,

$$\begin{pmatrix} E_{trans} \\ \bar{E}_2 \end{pmatrix} = T \begin{pmatrix} E_1 \\ E_{refl} \end{pmatrix}$$ (Eq. S3)

$$\begin{pmatrix} E_2 \\ \bar{E}_2 \end{pmatrix} = P \begin{pmatrix} E_1 \\ \bar{E}_1 \end{pmatrix}$$ (Eq. S4)

$$\begin{pmatrix} E_{trans} \\ \bar{E}_2 \end{pmatrix} = T_{end} \begin{pmatrix} E_1 \\ E_{refl} \end{pmatrix}$$ (Eq. S5)

where

$$T = \frac{1}{T_g} \begin{pmatrix} T_g^2 - R_g^2 & R_g \\ -R_g & 1 \end{pmatrix}$$ (Eq. S6)

$$P = \begin{pmatrix} e^{i(\beta+i\alpha)L_{net}} & 0 \\ 0 & e^{-i(\beta+i\alpha)L_{net}} \end{pmatrix} \tag{Eq. S7}$$

$$T_{end} = \frac{1}{T_e}\begin{pmatrix} T_e^2 - R_e^2 & R_e \\ -R_e & 1 \end{pmatrix} \tag{Eq. S8}$$

The elements of the input and output vectors are defined in Fig. S1(b). Fig. S1(c) sketches the $T$, $P$, and $T_{end}$ matrices. In Eq. S8, the transmission at the open end, $T_e$, is not well defined since the guided mode does not exist in free space. It turns out that $\Gamma$ is completely independent of $T_e$. The entire gap-rod termination is characterized by the total transfer matrix

$$M = T_{end}PT \tag{Eq. S9}$$

which connects incoming and reflected fields $E_1$, $E_{refl}$ on the left side with transmitted fields $E_{trans}$ on the right side of the termination

$$\begin{pmatrix} E_{trans} \\ \bar{E}_2 \end{pmatrix} = \begin{pmatrix} M_{11} & M_{12} \\ M_{21} & M_{22} \end{pmatrix}\begin{pmatrix} E_1 \\ E_{refl} \end{pmatrix}. \tag{Eq. S10}$$

Since we consider a single-channel input from the left side, $\bar{E}_2 \equiv 0$. Solving this system of equations yields the complex-valued total reflection

$$\Gamma = \frac{E_{refl}}{E_1} = -\frac{M_{21}}{M_{22}} = R_g + \frac{T_g^2}{R_g} \cdot \left(\frac{1}{1-R_eR_ge^{2(-\alpha+i\beta)L_{net}}} - 1\right), \tag{Eq. S11}$$

which is equivalent to that obtained via the power series ansatz (Eq. S2) discussed above.

In the following, we demonstrate the accuracy of the semi-analytical power series formula. For this purpose, in Lumerical MODE and FDTD solutions we determine the following parameter values (frequency = 361 THz, gap = 10 nm): $R_g = 0.7998\exp(i\,1.1495\,rad)$; $R_e = 0.9492\exp(i\,1.0886\,rad)$; $\alpha = 3.90533 \cdot 10^{-4}\,nm^{-1}$; $\beta = 0.016078\,nm^{-1}$; $T_g = 0.5579\exp(-i\,0.3313\,rad)$ and plug them into Eq. S2. Furthermore, we run full FDTD simulations of plasmonic wire-rod structures at 361 THz frequency and a constant gap of 10 nm, and compute the total reflectivity for various rod lengths. As seen in Fig. S2, the power series model perfectly reproduces the pronounced minima in reflection amplitude and the steep jumps in the reflection phase at the corresponding rod lengths as determined from FDTD simulations. The results from full FDTD simulations and those from the semi-analytical approach are in perfect agreement. This suggests that the radiation losses are well included in the simulated coefficients which enter the power series and transfer matrices. At the end open to vacuum, $|R_e| < 1$, and across the gap we obtain $|R_g|^2 + |T_g|^2 < 1$ for all gap widths and frequencies, *i.e.* a fraction of energy is systematically lost to free-space radiation when the plasmon propagation is perturbed by a discontinuity. Along the cylindrical sections of nanowire and nanorod, however, the only dissipation mechanism is Ohmic damping, characterized by $\alpha$.

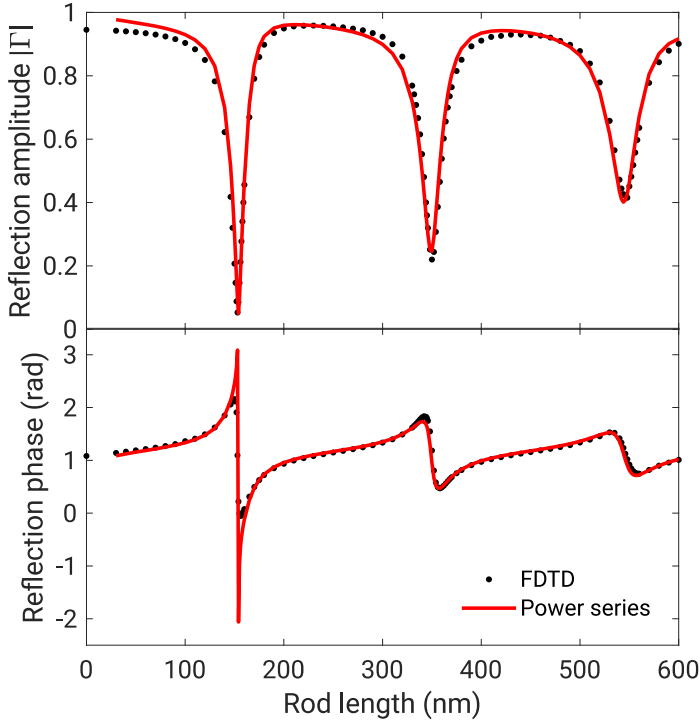

**Figure S2.** Reflection amplitude $|\Gamma|$ (upper panel) and phase (lower panel) of the wire-rod system. The total reflection is modeled by the interference of direct reflection at the gap and back-transmitted fields from the rod resonator. Black data points represent results from full FDTD simulations. The pronounced features in total reflection are fully captured by the analytic approach (Eq. S2, red solid line).

## III. Phase of the electric field on opposing sides of the gap at the "nonbonding" condition

At the gCPA condition, the guided plasmons on the nanowire exhibit zero reflectance at the nanorod termination. The rod lengths for gCPA and other reflection minima coincide with those of solitary resonant nanorods, as if the nanowire was absent. Here we show that this "nonbonding" condition exhibits a phase difference of $\pi/2$ between the electric field on opposing sides of the gap. We investigate the electric field at the rod termination for these gCPA scenarios based on the first and second order antenna resonances on the nano-rod. In both cases, the phase differences of the electric field and currents across the gap are close to $\pi/2$ (Fig. S3), as opposed to the bonding and antibonding resonances of a two-wire antenna, where the phase jumps of the field (current) over the gap amount to $\pi$ (0) and 0 ($\pi$), respectively. The $\pi/2$ phase shift ensures zero charge accumulation on the wire while the charge density is maximum at the rod ends, resulting in the absence of coupling between rod and feeding wire, *i.e.* a "nonbonding" condition. This nonbonding condition is fulfilled for all reflection minima observed in Figs. 2c-d of the main text, including the one where gCPA occurs.

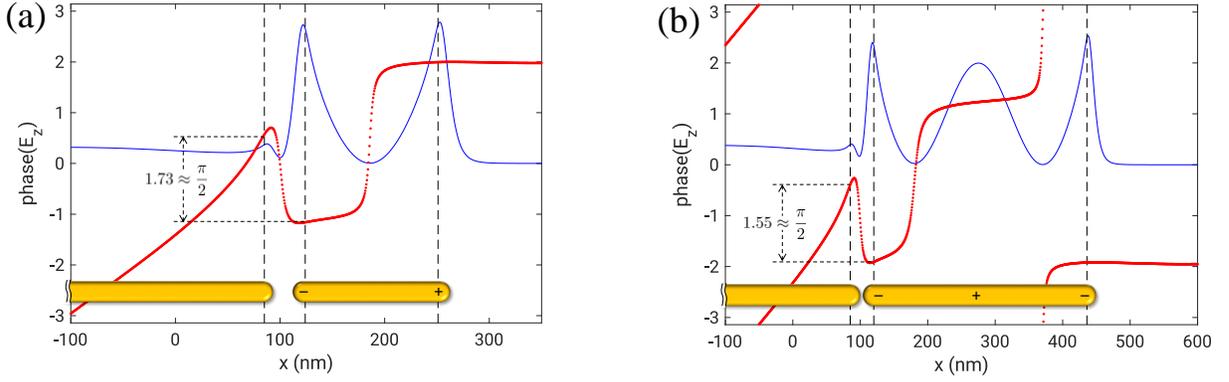

**Figure S3.** Phase of the *z*-component of the electric field (red dots), recorded 5 nm away from the wire surface for (a) the dipolar resonance on a 157 nm long nanorod separated by a 9 nm gap at 354.5 THz and (b) the second-order antenna resonance on a 346 nm long nanorod separated by a 5 nm gap at 362.2 THz. The phase jump across the gap is close to $\pi/2$ in both cases. The blue solid line depicts the total field intensity. The symmetry of charges is indicated on the structure outlines.

## IV. The relation between the propagation constant $\beta$ and the damping $\alpha$ of the guided $TM_0$ mode.

By means of a frequency-domain eigenmode solver (MODE Solutions, Lumerical Solutions Inc.) we determine the dispersion relation of the fundamental guided $TM_0$ mode. This yields a connection between the allowed values for propagation constant $\beta$, attenuation constant $\alpha$, and frequency (Figs. S4(a)-(b)). For each frequency point, a pair of $(\alpha, \beta)$-values represents the dispersive characteristic of the mode, with which we are able to define a relation $\alpha(\beta)$ that describes the $TM_0$ mode as a line on the complex wavevector plane (Fig. S4(c)) by eliminating the explicit frequency dependence. The same line is depicted in white in Fig. 3a and in red in Fig. 3b of the main text. Only points lying on this curve in the *k*-plane are physically accessible by the $TM_0$ mode.

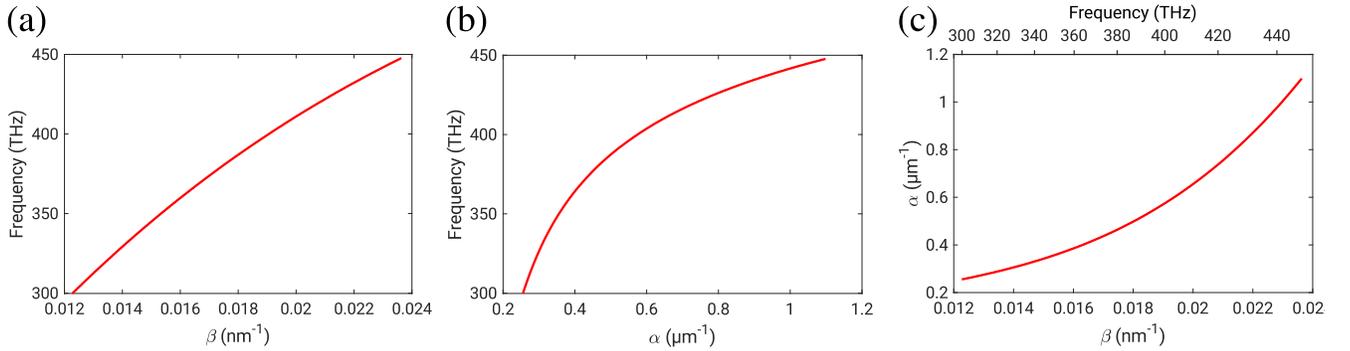

**Figure S4.** The complex dispersion relation of the $TM_0$ mode is given by (a) the propagation constant $\beta$ and (b) the attenuation constant $\alpha$ as a function of frequency where $k = \beta + i\alpha$. The wavevector data are drawn in (c) as a line within the complex *k*-plane.

## V. Animation of the moving phase singularities

This movie is available online.

Description:
The movie contains the reflection phase in the upper panel and the spectrum of the reflected intensity in the lower panel. At the beginning of the movie, the geometry of the wire-rod system is not optimized for gCPA. Therefore, the phase singularity is located far away from the modal line of allowed propagation and attenuation constants of the guided $TM_0$ mode (white solid line in the movie) and the reflectivity is far from zero. In the first half of the movie, the rod length is gradually decreased from 160 nm to 140 nm with the gap size fixed at 4.0 nm. The phase singularity moves horizontally in the complex $k$-plane towards higher $\beta$-values, in accordance with the shift of the dip in the reflectance spectrum towards higher frequencies. The minimum reflection is still at ~10%.

Then the nanorod length is kept at 140 nm and the gap size is gradually enlarged from 4.0 nm to 30.0 nm. This gradually decreases the reflectance. At gap = 9.0 nm (around 6s in the movie), the phase singularity resides on the modal line of the accessible values of $\alpha(\beta)$, leading to zero reflection at 381 THz due to gCPA. After crossing the modal line the singularity continues its motion toward lower $\alpha$-values which leads to increased reflectance.

A further interesting aspect is the second phase singularity, located in the negative $\alpha$-region of the wavevector plane. It is associated with a reflection pole of the wire-rod structure (see Fig. 3a of the main text) and would be physically observable if gain was added to the nanorod. As the gap increases in the movie, this phase singularity and the pole slightly approach the real axis of the $k$-plane, $i.e.$ they flow towards lower gain values. This behavior might facilitate the observation of a low-threshold SPASER operation.

## VI. Two-port generalized CPA for coherent control of nanoantenna radiation

The appearance of CPA in dielectric resonators with two input ports was demonstrated earlier [1,2]. S-matrix approaches were used to model the output intensity and to find zeros therein. Likewise, for a plasmonic wire-rod-wire system (Fig. S5(a)) consisting of two semi-infinite wires sandwiching one nanorod by two identical gaps, a total transfer matrix of the form

$$M = TPT \tag{Eq. S12}$$

can be calculated (definitions of $P$, $T$ see section II). Generalizing the concept of two-port CPA according to section 2 of the main text, the subset of incoming and outgoing guided plasmons is investigated. Converting $M$ into a scattering matrix $S$ defined by the basis vectors

$$\begin{pmatrix} E_{out,l} \\ E_{out,r} \end{pmatrix} = S \begin{pmatrix} E_{in,l} \\ E_{in,r} \end{pmatrix} \tag{Eq. S13}$$

requires a basis transformation of the form

$$S = \begin{pmatrix} S_{11} & S_{12} \\ S_{21} & S_{22} \end{pmatrix} = \begin{pmatrix} -M_{21}/M_{22} & 1/M_{22} \\ M_{11} - M_{12}M_{21}/M_{22} & M_{12}/M_{22} \end{pmatrix}. \tag{Eq. S14}$$

For a two-port excitation of equal intensities and tunable relative phase $\phi_{in}$, the input vector reads

$$\begin{pmatrix} E_{in,l} \\ E_{in,r} \end{pmatrix} = \begin{pmatrix} 1 \\ e^{i\phi_{in}} \end{pmatrix} \tag{Eq. S15}$$

The output intensity obtained from Eqs. S13-15 is defined as

$$I_{out} = \frac{|E_{out,l}|^2 + |E_{out,r}|^2}{|E_{in,l}|^2 + |E_{in,r}|^2} = \frac{|E_{out,l}|^2 + |E_{out,r}|^2}{2} \tag{Eq. S16}$$

Zero eigenvalues of $S$, *i.e.* zero output intensity is found for specific sets of rod length, gap size, and frequency, either for symmetric input ($\phi_{in} = 0$) or antisymmetric input ($\phi_{in} = \pi$).

In Fig. S5(b) we show the simulated electric field patterns 5 nm above a wire-rod-wire system with rod length 154 nm, gap size 23 nm on both sides at a frequency of 360.9 THz. The $z$-component field intensity is displayed for symmetric and antisymmetric input. Fig. S5(c) provides maps of the total electric field intensity in the vicinity of the nanorod. With symmetric input ($\phi_{in} = 0$), the output waves interfere constructively, leading to a pronounced standing wave along the two wires ("superscattering"). At the same time, the fields around the nanorod are highly suppressed. Therefore, the energy can hardly enter the nanorod and the antenna is "off". In contrast, with antisymmetric input ($\phi_{in} = \pi$), the interference of the transmission and reflection is perfectly destructive, leading to gCPA of the input power by the nanorod and highly enhanced field intensity on the nanorod. In this case, the antenna is "turned on" and perfectly driven by the two wires.

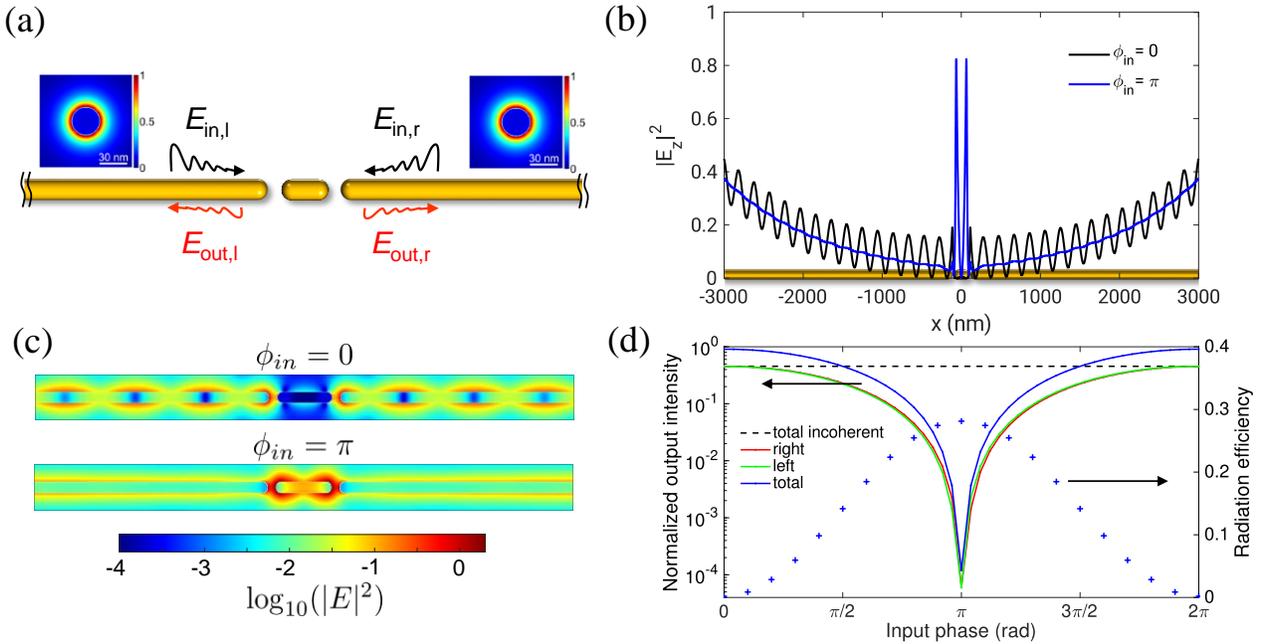

**Figure S5.** (a) Sketch of the symmetric wire-rod-wire system where two-port gCPA is realized, *i.e.* $E_{out}$ vanishes on both sides. Insets show field distributions of the guided TM$_0$ mode. (b) Simulated intensity of the main electric field component ($E_z$) along the structure, 5 nm above the surface of the wire. The black line shows the case of symmetric input ($\phi_{in} = 0$) with constructive interference of input and output waves. The antisymmetric input case ($\phi_{in} = \pi$) is plotted as a blue line. This yields gCPA on the nanorod and destructive interference along the wires. (Rod length 154 nm, gap width 23 nm, frequency 360.9 THz) (c) Corresponding maps of the total field intensity in logarithmic color scale. (d) Calculated normalized output intensities on the left side (green), right side (red) and the sum of both (blue) under coherent input. The simulated radiation efficiency of the nanorod as a function of input phase $\phi_{in}$ is plotted with blue plus signs (scale to the right). With an input phase of $\pi$, we achieve gCPA with output intensities on the wires as low as $10^{-4}$, accompanied by maximum far-field radiation from the dipolar mode on the nanorod. The horizontal black dashed line depicts the incoherent sum of the squared transmission amplitudes on left and right side.

The input phase $\phi_{in}$ constitutes an additional degree of freedom within the two-port structure, compared to the wire-rod system. It is an external parameter that can be adjusted *e.g.* by interferometric pathways, thereby enabling coherent control of light by light without using any nonlinear effect. Generalized CPA, superscattering, and any absorption state in between can be reached by tuning $\phi_{in}$. At CPA, the output intensity of the wire plasmons can be suppressed by four orders of magnitude (Eq. S16; Fig. S5(d)). In the all-plasmonic system studied here, a further aspect comes into play. Making use of a radiative dipolar antenna mode, the control of the input phase $\phi_{in}$ allows the control of conversion of the input power into far-field radiation. Fig. S5(d) shows the radiation efficiency of the nanorod at different input phase $\phi_{in}$. At $\phi_{in} = 0$, the rod is in an "off" state and no radiation is observed. When $\phi_{in}$ approaches $\pi$, the radiation increases up to 30% of the overall injected power. As a benefit of coherence in the absorption process, a nonlinear mechanism is not required for such switching and deep modulation.

# VII. References

[1] Y. D. Chong, L. Ge, H. Cao, and A. D. Stone, "Coherent perfect absorbers: time-reversed lasers," Phys. Rev. Lett. **105**, 053901 (2010)

[2] W. Wan, Y. Chong, L. Ge, H. Noh, A. D. Stone, and H. Cao, "Time-reversed Lasing and Interferometric Control of Absorption," Science **331**, 889-892 (2011)